\documentclass[10pt,a4paper]{article}
\usepackage[latin1] {inputenc}
\usepackage{authblk}
\usepackage[cmex10]{amsmath}
\usepackage{amsfonts,amssymb,amsthm}
\usepackage{algorithmic,algorithm}

\usepackage{color,soul}

\newtheorem{definition}{Definition}

\newtheorem{lemma}[definition]{Lemma}

\newtheorem{claim}{Claim}
\newtheorem{theorem}[definition]{Theorem}

\usepackage{enumitem}

\newcommand{\gossip}{$\mathcal{GOSSIP}$}
\newcommand{\local}{$\mathcal{LOCAL}$}

\newcommand{\CE}{{\mbox{CE}}}

\newcommand{\prot}{{\mathcal{P}}}
\newcommand{\good}{{\mathcal{G}}}
\newcommand{\Exp}{{\mathbf{E}}}
\newcommand{\Prob}{{\mathbf{Pr}}}
\newcommand{\ffaults}{\alpha}

\newcommand{\plenght}{\gamma}

\newcommand{\Expec}[2]{\mathbf{E}_{#1} \left[ #2 \right]}

\newcommand{\equilibrium}{whp\,$t$-strong equilibrium}

\newcommand{\intention}{{\textsc{Voting-Intention}}}
\newcommand{\asking}{{\textsc{Commitment}}}
\newcommand{\voting}{{\textsc{Voting}}}
\newcommand{\findmin}{{\textsc{Find-Min}}}

\newcommand{\verification}{{\textsc{Verification}}}
\newcommand{\coherence}{{\textsc{Coherence}}}

\newcommand{\sA}{ {\mathcal A}}

 \newcommand{\skproof}{\noindent\textit{Sketch of Proof. }}
  \newcommand{\bigproof}{\noindent\textit{Proof. }}

\begin{document}

\title{\textbf{ Rational Fair Consensus in the \gossip\ Model} \\ }

\author[1]{Andrea Clementi}
\author[1]{Luciano Gual\`a}
\author[2]{Guido Proietti}
\author[2]{Giacomo Scornavacca}
\affil[1]{Universit\`a \emph{Tor Vergata} di Roma, {\tt clementi/guala@mat.uniroma2.it}}
\affil[2]{  Universit\`a degli Studi dell'Aquila, {\tt guido.proietti@univaq.it\\ giacomo.scornavacca@graduate.univaq.it  }}

\maketitle

\begin{abstract}
The \emph{rational fair consensus problem} can be informally defined as follows.
Consider a  network of $n$ (selfish) \emph{rational agents}, each of them initially supporting a   \emph{color}  chosen  from a finite set $ \Sigma$.
The goal is to design a protocol that  leads the network  to   a stable monochromatic configuration (i.e. a consensus)  such that the probability that
the winning color  is $c$ is equal to the fraction of the agents that  initially support $c$, for any $c \in \Sigma$.
Furthermore, this fairness property must be guaranteed  (with high probability) even in presence of
  any fixed  \emph{coalition} of rational agents that may deviate from
the protocol in order to increase the winning probability of  their supported colors.  A protocol  having  this  property,
 in presence of coalitions of size at most $t$, is said to be a \emph{whp\,-$t$-strong  equilibrium}.

We   investigate, for the first time, the rational fair consensus problem in the \gossip\ communication model where, at every
round, every agent can actively contact at most  one neighbor   via a   \emph{push$/$pull} operation.
We provide a randomized  \gossip\ protocol that, starting from any
 initial color configuration of the complete graph, achieves
  rational fair consensus within $O(\log n)$ rounds using messages of $O(\log^2n)$ size, w.h.p.
More in details, we prove that our  protocol   is     a \equilibrium\   for any $t = o(n/\log n)$ and, moreover,
 it tolerates worst-case permanent faults provided that the number of non-faulty  agents is $\Omega(n)$. As far as we know, our protocol is the first solution which avoids any all-to-all communication, thus resulting in  $o(n^2)$ message complexity.

\end{abstract}

\section{Introduction}

There is an increasing interest on algorithmic  tasks performed  by distributed systems that are
formed by a finite set of   selfish,  \emph{rational agents}.
When the system does not provide for any central authority, the techniques for studying
  this kind of processes     lie  at the intersection of two scientific fields,  \emph{Distributed Computing}
 and    \emph{Algorithmic Game Theory}.
Typically, there is a social task/decision to be performed by the distributed
system  and, at the same time, every agent of the system has his own profit: The latter being  a fixed
function of the   final configuration reached by the system. A feasible solution  here  consists in a protocol that not only computes  the desired task but it  even must result profitable for
every rational agent: In other words, agents should not get  any gain (according to their own profit functions) to  deviate from the
protocol's local rules. Protocols satisfying this property are said \emph{Nash equilibria}  or, even better, \emph{$t$-strong (Nash) equilibria} when this robustness property is guaranteed even if agents can form a deviating \emph{coalition} of size at most $t$ \cite{AFM07}.
 Perhaps,    this framework  has been investigated for the first time in   \cite{HT04},  where
 rational behaviour is analyzed in  secret-sharing problems and multiparty computations. More recently, the impact  of rational adversaries  has been investigated
on    fundamental tasks in Distributed Computing such as \emph{leader election}, \emph{consensus},
 and \emph{wake-up}  problems \cite{AbrahamDH13,AfekGFS14,GroceKTZ12,HV16}.

Inspired by this new line of research, we here study the \emph{rational fair consensus}  problem \cite{HV16} which can be informally defined as follows (see Section \ref{sec:prely} for a formal definition). At the onset, every agent supports
a \emph{color} $c \in \Sigma$ where $\Sigma$ is the   color space.
The goal  of the system is  to reach a stable
monochromatic  configuration where all agents support the same \emph{winning} color  and the probability that the winning color is $c$ equals the fraction of agents initially supporting $c$, for any color $c \in \Sigma$.\footnote{Notice that this \textit{fairness} property is stronger than validity, the latter being required in the classic consensus problem - see Section \ref{sec:prely}.} Moreover, every rational agent $u$ has his own profit function which is maximized
whenever the winning color is the one he supported at the onset, while it is smaller when the winning color is any other color, and, finally,  it  is much smaller (and minimized) whenever the protocol fails to achieve consensus.
More general  classes of profit functions have been studied for rational fair consensus (e.g.
 \cite{AbrahamDH13}). The well-known
\emph{fair leader election problem} is the special case of the fair consensus problem where the color initially supported by
each agent   is  his own ID.

In this setting, a \emph{with high probability}\footnote{We will adopt here the standard notion in
probabilistic algorithms: An event is said to hold with high probability if its probability is
$1- \frac{1}{n^{\Omega(1)}}$.} (for short, \emph{w.h.p.}) $t$-strong  equilibrium  (see Def. \ref{tstrong})  for the rational fair consensus problem is a protocol $\prot$ that, given any initial color configuration $\vec c$ for the $n$ agents,
w.h.p.  achieves fair consensus and, moreover, for any coalition
$C$ of size at most $t$ and for any deviating strategy of $C$, there is at least one  agent in   $C$ that, according to
the deviating strategy,  w.h.p.\ will not  increase his   chance to make  his  color win.

Several versions of consensus
 in presence of rational agents have been recently studied \cite{AbrahamDH13,AfekGFS14,GroceKTZ12} but only few of them consider the  fairness property.
 As far as we know, rational   consensus has been  studied only in the \local\ communication model \cite{HHKM12,HP02} where, at every round,
 every agent can exchange messages with all his neighbors.  In    \cite{AbrahamDH13},   Abraham et al.
 present a  protocol for fair leader election that  is an \emph{$(n-1)$-resilient equilibrium} ($t$-resilient equilibrium is a stronger version
   than $t$-strong equilibrium, where no agent of the coalition will profit from a deviation \cite{AbrahamDGH06}).   However, their protocol
 is not robust against crash faults.
 A   protocol achieving   consensus  is given in \cite{GroceKTZ12} in presence of a rational adversary that controls a proper subset of agents (this model is different  from the ones studied in  \cite{AbrahamDH13,AfekGFS14,HV16} and in this work).
 Their protocol does not guarantee fairness and assumes there are no crash faults.  In \cite{AfekGFS14},  a protocol achieving
   $(n-1)$-resilient equilibrium\footnote{Actually, in \cite{AfekGFS14}, the obtained property is improperly named as $t$-strong equilibrium.}   is provided   for fair leader election, while, in presence of crash faults, the protocol is shown to be (only) a Nash equilibrium. This protocol does not work for the rational fair consensus problem and, in general,  we emphasize that,   even though any protocol for
      fair leader election  can be easily transformed to one for   fair consensus,  if agent's rational  behaviour is   considered - i.e.
     if  the ``rational'' versions of such two problems are considered -,
      then this reduction  is no longer true \cite{HV16}.
   Further results are presented  in   \cite{BeiCZ12} for  some   versions of   rational consensus  in models which
     depart significantly from ours.

  More recently, Halpern and Vilaca \cite{HV16}
 studied the rational fair consensus problem in the   \local\ communication model  assuming that agents  have unique IDs.
  They study  the problem in the complete graph and in presence
    of   dynamic patterns of crash faults.  It is first  shown that if the adversary
  can adaptively choose the initial configuration and the dynamic fault-pattern (i.e. a worst-case, dynamic adversary), then no protocol can achieve a Nash equilibrium  for  rational fair consensus.
   Then, they consider a much weaker, natural adversary on the complete graph: The adversarial fault pattern is chosen randomly according to some distribution $\pi$. They prove that, if $\pi$ satisfies some reasonable conditions, then  it is possibile to design a protocol achieving a Nash equilibrium for rational fair consensus provided that the overall number of faulty agents  is smaller than $n-1$. Their protocol  is not  ``light-weight''  \cite{HP02} since it has to take care about  random  dynamic fault patterns and it requires
    $\Omega(n^2)$ messages. The authors also claim that  its robustness against coalitions
     is quite   hard to   analyze and thus it is   not included in the paper.

  \smallskip
  \noindent
  \textbf{Our contribution.} Recently, there has been strong interest in the design of algorithms for several versions of consensus problems  in network models that severely restrict communication and computation \cite{AAE07,BCNPS15,DGMSS11}: This both for efficiency considerations and because such models capture aspects of the way consensus is reached in social
networks, biological systems, and other domains of interest in network science. From the point of view of computation, the restrictive setting is to assume that each node only has polylogarithmic size of memory available, while, as for communication, this bound is also required to link bandwidth available in each round. Finally, the number of interactions a node can open in one round are severely constrained.
These constraints are well-captured by the synchronous \gossip\ model \cite{BCNPS15,DGHILSSST87,DGMSS11,HHKM12,KSSV00,KDG03}: At every round, 
every node can actively push or pull  a (short) message (say, of polylogarithmic size) with at most one of his  neighbors. Notice that, in every round,
 a node can receive more than one message     
but   the   number of active links   is always $O(n)$:  A
  \emph{per-round} communication pattern that can be considered definitely  reasonable
  in  several real network applications.

A major point is that all the  previous  protocols    for rational (fair) consensus \cite{AbrahamDH13,AfekGFS14,HV16}
   heavily rely on broadcast operations made by every (non faulty) agent: In the complete graph, every agent directly communicates some piece of information (e.g. his own ID) to all the agents.   It turns out that  the  number of exchanged  messages
   is $\Omega(n^2)$.

       Achieving Nash equilibria without the use of all-to-all operations is a major technical
         issue  we want to investigate in this paper.

  As in the work of Halpern and Vilaca  \cite{HV16}, we consider a complete network of $n$ agents, each of them
  having a unique ID which is an integer in the set $[n] := \{ 1 , \ldots, n\}$.
  We consider the \gossip\ communication  model and, concerning   rational fair consensus,
  we assume each agent $u$ initially  knows his ID, his initial opinion $c_u \in \Sigma$ and the network size $n$.
  When two nodes communicate, despite their selfish behaviour, they cannot cheat each other about their ID's.
   This condition is more than reasonable in several network scenarios, it is assumed in the previous works (e.g. \cite{HV16}) and it definitely does not make the problem easy.
  Taking in mind the strong negative result obtained by Halpern and Vilaca  \cite{HV16} about worst-case dynamic
  agent faults, we explore a weaker kind of adversary, the \emph{worst-case permanent} one: At the very beginning, every agent
  can be either  active or faulty and we assume  this initial setting can be    managed by a worst-case adversary.
  After this setting, then no further action of the adversary is allowed. We again remind  that
    rational fair consensus in any   model allowing only sparse  communication patterns has never been studied
   so far, even in the fault-free case: This communication constraint essentially
   makes previous solutions of little use. Moreover, the presence of this  static adversary    introduces  further
   issues to take care about: a rational active agent can pretend to be a faulty node in some rounds, and hence the protocol must be robust also against this kind of (potentially profitable) deviations.

 We provide a \gossip\ protocol that, starting from any
 initial color configuration, achieves rational fair consensus in $O(\log n)$ rounds using local memory and messages of $O(\log^2n)$ size, w.h.p., thus resulting in $O(n \log ^3 n)$ overall communication complexity.
 We prove that our  protocol   is  a whp\,-$t$-strong  equilibrium   for any $t = o(n/\log n)$ and, moreover,
 it tolerates worst-case permanent faults provided that the number of active agents is $\Omega(n)$.
  We   remark that
the known previous protocols   \cite{AbrahamDH13,AfekGFS14,HV16}, on the complete graph, use  $\Omega(n^2)$   messages and local memory of size $\Omega(n)$.   It is always possible to simulate a \local\ protocol over the \gossip\ model thanks to the general technique introduced  in \cite{HHKM12}. However,
 this approach  would yield   exponentially larger message size and it is not clear whether
the so-obtained simulation achieves any kind of equilibrium  w.r.t.    selfish behaviour.

 To the best of our knowledge, our  protocol is  the first efficient solution for rational fair consensus
 on the \gossip\ model and, thus, it represents a first evidence of the fact that  a short
  sequence of    sparse communications patterns (each pattern formed by $n$ push/pull operations) suffices
  to reach this kind of   equilibria. We believe this result might open interesting directions in the design of
  more scalable solutions in real network applications where fair consensus in presence of selfish agents
   is a crucial issue  \cite{bit_crypto}.

\section{Preliminares} \label{sec:prely}

We consider a complete graph $G([n],E)$ of $n$ nodes, each of them  having a unique \emph{label} in $[n] = \{1,\ldots , n\}$,
 and we adopt the synchronous \gossip\  model: at every round, each node can make either a \emph{pull} or a \emph{push} operation with one of his neighbors. The choice of the neighbor can be made \textit{uniformly at random} (for short, u.a.r.). At the onset, every node $u$ knows $n$ and how to communicate with every other node over a \emph{secure channel}: during a   communication over the edge $\{u,v\}$,  the  two nodes are aware about the label of his  peer and the exchanged message is private (this is fully in the line of  the related
 previous works \cite{AbrahamDH13,AfekGFS14} and, moreover, it well reflects the real scenarios inspiring the \gossip\ model, such as peer-to-peer and opportunistic networks).

   The classic  \emph{consensus problem} in presence of \emph{unknown, permanent node-faults} can be defined as follows. 
    At round $t=0$, every node is either in  the \emph{active} state or in the \emph{faulty} state and let $\sA$ be the subset of active nodes. The permanent faults are chosen by a worst case adversary that knows the protocol. A node, starting  in the faulty state,
     will remain quiescent for all the process while each active node $u \in \sA$   supports a color $c_u \in \Sigma$ ($\Sigma$ being a shared set of colors). 
     A protocol solves the  \emph{consensus task} if all the following conditions are met:

\begin{itemize}[leftmargin=*]
\item \emph{Termination}: Every active node gets into a final state within a finite number of rounds.
\item \emph{Agreement}: When all active nodes have reached a final state they will support the same color $c$. We say that $c$ is the winning color.
\item \emph{Validity}: The winning color $c$ must be a \emph{valid} one, i.e.,
a color which was initially supported  by at least one active node.
\end{itemize}

\noindent
  In the \emph{fair-consensus} task \cite{HP02,HV16}  the  \emph{validity} property
   is replaced by a stronger, probabilistic property.

\begin{itemize}[leftmargin=*]
\item \emph{Fairness}: The probability  that a color $c \in \Sigma$ is the winning one is  equal to the fraction of active nodes that initially support  $c$.
\end{itemize}

We remark that, initially,  every node only knows  his label and his state (active or not) while he  knows nothing
 about the other nodes. It is only assumed that  the (unknown)  set $\sA$   has   linear size, i.e. $\vert \sA \vert = \Theta(n)$.
 A well-studied special case of the above task is the \emph{fair leader  election} where every   node initially support his own  ID as a color and, hence, every active node must have the same chance to be elected.

\smallskip
\noindent
\textbf{Non-cooperative setting.}
Besides permanent node faults, we consider nodes that act as  \emph{selfish  (rational) agents}  according to the standard definition in Game Theory.
 For this reason, we denote the problem as  the \emph{rational fair consensus}.  Formally, each
  node (from now on, \emph{agent}) $u \in [n]$
 has a utility function $\mathrm{util}_u(s)$ defined  on every  final state $s \in S$ of the protocol,
  where $S = \Sigma \cup \lbrace \perp \rbrace$ (the protocol can either converge to a color or, if agreement is not reached, fail).
  We focus on the  natural scenario  where    the utility function of agent $u$ is maximal when the winning color is $c_u$,
 it  is much less when the protocol converges to  another color, and, it is minimal (in fact, it is very bad) when the protocol fails. We assume the following (normalized) payoff scheme: For each agent $u$ there is exactly one value $c_u \in \Sigma$ such that $\mathrm{util}_u( c_u) = 1$; moreover,  $\mathrm{util}_u(s = \perp) = -\chi$, for an arbitrary fixed value $\chi \geq 0$, while  $\mathrm{util}_u( c') = 0$ otherwise.

The strategy of an agent $u$ is that of choosing an \textit{adaptive}   local
algorithm $\sigma_u$ from a set  of feasible
rules satisfying the system constraints. The adaptive algorithm defines the actions  of an
 agent at every  round: These actions may depend on the set of messages
  received  so far during the process.
  Each agent chooses such an algorithm in order to maximize his expected utility,
  where the expectation is defined over the random choices performed by the agents during the process.
  A protocol thus results in the vector of the $n$ local (randomized) algorithms chosen by every agent (also called \emph{strategy profile} in Game Theory). Given a protocol $\prot$ and an initial color configuration $\vec c$, we call
  $Q(\prot , \vec c)$ the set of all the possible executions of $\prot$ starting from $\vec c$.
Moreover, let $q(\prot , \vec c)$ be  the random variable over $Q(\prot , \vec c)$ representing a random execution of $\prot$. We define
$f : Q(\prot , \vec c) \rightarrow S$ as the function that  returns the outcome of any execution.
For  brevity's sake, for any agent $u$,  we define  $r_u( \prot , \vec c) = \mathrm{util}_u(f(q(\prot , \vec c)))$.

We adopt the following notion of equilibrium, called \emph{\equilibrium} that it is a probabilistic relaxation of $t$-strong equilibrium (a similar relaxation is considered in \cite{GH05} for deterministic \emph{truthfulness}).  Such an equilibrium is a protocol (strategy profile)  such  that,  for any deviation of any fixed coalition of size at most $t$, there is an agent in the coalition that will not improve his expected utility, w.h.p. This is formalized by conditioning the expected utility of the agents to a large subset of ``good'' executions of the protocol. Formally, let us consider a
 protocol $\prot$, a coalition $C \subseteq [n]$ and a (restricted) protocol $\prot'$ for $C$. By
 $(\prot_{-C},\prot'_{C})$ we denote the protocol where the agents in $\sA \setminus C$ follow $\prot$ while
 the active agents in $C$ follow $\prot'$. Given a color configuration $\vec c$, protocols $\prot$ and $\prot'$, we
 let  $\Omega = \Omega(\prot , \vec c)$ and  $\Omega' =
  \Omega((\prot_{-C},\prot'_{C}), \vec c)$ be the probability spaces yielded by running $\prot$ and $(\prot_{-C},\prot'_{C})$
  from $\vec c$, respectively.

\begin{definition}\label{tstrong}
We say a protocol $\prot$ is a \emph{whp\,-\,$t$-strong equilibrium}  if, for any initial color configuration
$\vec c$,  for every coalition $C$ of at most $t$ agents, and for every restricted protocol  $\prot'$ for $C$,
the following properties hold:

\begin{itemize}[leftmargin=*]
\item There is a subset $\good \subseteq \Omega$ of executions such that
$\Pr_{\Omega}(\good) \geq 1 - \frac 1{n^{\Theta(1)}}$;

\item There is a subset $\good' \subseteq \Omega'$ of executions such that
$\Pr_{\Omega'}(\good') \geq 1 - \frac 1{n^{\Theta(1)}}$;

\item There is an agent $w \in C$ such that \begin{align}\begin{split}
 \Exp_{\Omega}&[r_w( \prot , \vec c) \vert q((\prot , \vec c) \in \good] \geq \\
 \Exp_{\Omega'}&[r_w( (\prot_{-C},\prot'_{C}) , \vec c) \vert q((\prot_{-C},\prot'_{C}), \vec c) \in \good'].
\end{split}
\end{align}
\end{itemize}
\end{definition} 
\section{An Efficient Protocol for Rational Fair Consensus} 

\begin{algorithm}
\begin{algorithmic}
\STATE {\large \textbf{The local  rules for Protocol $\prot$}}

\textbf{Local Data:} Each agent $u \in [n]$ knows     label $u$,  his supported color $c_u \in \Sigma$,  the agent number  $n$, and the fault-tolerance parameter $\plenght$;\\
\textbf{\textsc{Initialize()}}: $u$ computes the parameters $m = n^3$ and the number of rounds $q = \plenght \log n $;

\textbf{\intention}(): Choose a list of votes $H_u$ \\
$H_u : = \lbrace (h_{u,1},z_{u,1}) \dots (h_{u,q},z_{u,q}) \rbrace$ where \\ $\forall i$ in $[q]$ $h_{u,i}$ is chosen u.a.r. in $[m]$ and $z_{u,i}$ is chosen u.a.r. in $[n]$\;

\textbf{\asking}(): Compute a  list $L_u$ of collected vote intentions \\  
$L_u : = \emptyset$\;
\FOR{$q$ rounds}
	\STATE \textit{Pull} from an agent $v$ chosen u.a.r. his list $H_v$\footnotemark\;
	\STATE $\forall j \in [q]$ update $L_u : = L_u \cup \{ (v,h_{v,j},z_{v,j}) \}$; \\ 
	\STATE \textit{Receiving} a pull requests: send  your own  $H_u$ list\;
\ENDFOR

\textbf{\voting}($H_u$): Push your votes according to $H_u$ and collect the received votes in $W_u$ \\
$W_u : = \emptyset$\;
\FOR{$i = 1, \ldots , q$ rounds }
	\STATE \textit{Push}  $h_{u,i}$ to agent $z_{u,i}$\;
	\STATE \textit{Receiving votes:}  let  $\{h_1,  \ldots,   h_{\ell} \}$ be the votes received (in round $i$)	from   agents $\{z_1, \ldots , z_{\ell}\}$, respectively  and  update $W_u : = W_u \cup \{(h_1,z_1) \ldots (h_{\ell},z_{\ell})\}$\;
\ENDFOR

Compute the value $k_u : = \sum_{h \in W_u} h \mod m$\;

\textbf{\findmin}($\CE_u = (k_u,W_u,c_u,u)$):\\
$\CE^{min}_{u} : = \CE_u$\;
\FOR{$q$ rounds}
	\STATE  \textit{Pull} from an agent $v$ u.a.r. in $[n]$ his Certificate $\CE^{min}_{v}$\;
	\IF{ $k^{min}_v < k^{min}_u$}
		\STATE $\CE^{min}_u : = \CE^{min}_v$
	\ENDIF
	\STATE \textit{Receiving} a pull request: send $\CE^{min}_{u}$
\ENDFOR

- \textbf{\coherence} ($\CE^{min}_{u}$):\\
\FOR {$q$ rounds}
	\STATE \textit{Push} to an agent $v$ u.a.r. in $[n]$ the Certificate $\CE^{min}_{u}$\;
	\STATE \textit{Receiving} a set of \emph{Certificates} $\CE$:\\
	\IF{$\exists \CE^{min}_{v} \in \CE: \CE^{min}_{u} \neq \CE^{min}_{v}$}
	\STATE Make the protocol fail;
	\ENDIF
\ENDFOR

\STATE $\CE^{min} : = \CE^{min}_{u}$\;

\textbf{\verification}($L_u$): 
\STATE $\CE^{min}: =(k_{min},W_{min},c_{min},z_{min})$:\\
\IF {$k_{min} = (\sum_{h \in W_{min}} h \mod m)$ and $W_{min}$ is consistent\footnotemark\ with the list of votes in $L_u$}
		\STATE Support the color $c_{min}$
\ELSIF{} \STATE Make the protocol fail;
\ENDIF

\label{alg}
\end{algorithmic}
\caption{Protocol $\prot$}
\end{algorithm}

\footnotetext[4]{if agent $v$ does not reply (or replies in a unexpected way), then he is marked as faulty ($\forall j \in [q], h_{v,j} = 0$).}
\footnotetext[5]{For each $z_v$ appearing in $W_{\min}\cap L_u$ check if the vote is the same.}
\noindent
\textbf{Informal description of the protocol.}
In order to  reach \emph{fair consensus}, our protocol adopts a simple and natural  idea (see, for example \cite{AbrahamDH13}):
Choose u.a.r.  an active agent of  the network and then lead the system to stabilize on  the color supported by this agent.
  It is easy to show that if all the agents follow the protocol then a fair consensus is achieved. However,
  the presence of a coalition of rational agents
  requires further protocol actions in order to prevent convergence towards unfair consensus: This is obtained using some verification procedures that work in logarithimic time and use messages of size $O(\log^2n)$.


 The   protocol is parametrized in the maximum number $\alpha n$ of faulty agents (where $0 \leq  \alpha <1$ is the
so-called fault-tolerance parameter of the protocol)  and it
 assumes every agent knows  the size $n$ of the system.
 Its  local rules are organized    in the following consecutive   phases. A detailed description
 of the protocol is given in Algorithm 1.

 \noindent
 - In the \intention\ phase each agent $u$ randomly chooses   a ``small'' (i.e. a logarithmic)  number of    agents and, for each of them,
he  decides one  random \emph{vote}  (chosen u.a.r. in the range $[n^3]$): The resulting  list is called the \emph{vote intention} $H_u$ of agent $u$.

\noindent
- In the \asking\ phase, each agent $u$ asks (using  pull operations) a small number of agents to send him
their \emph{vote intentions}: All such data will be stored in  a set we  call  $L_u$. If an agent $v$ does not answer to one of $u$'s requests, then  $v$
 is marked as \emph{faulty} by $u$ and, from now on,  $u$ will consider  all the votes  of    $H_v$   equal to zero.

 \noindent
 - In the \voting\ phase, each  agent $u$
 votes (via the push mechanism) according to $H_u$ and, thus,  in turn,  $u$ also gets  the set   $W_u$ of the received votes from the other agents.
 Now   each agent $u$ can compute the value $k_u$ equal to the sum of all the received votes modulo $m = n^3$ and creates his \emph{Certificate} $\CE_u$. The certificate contains the value $k_u$, the received votes $W_u$, his color $c_u$ and his label $u$. The choice of this value for $m$ 
 ensures that all  $k_u$'s are different, w.h.p. and, so, the minimum is unique (this fact will be exploited in the next phase).  
 
 \noindent
 - All the agents start the  \findmin\ phase that makes
 every active  agent converge on  the ``minimal'' certificate $\CE_{z}=(k_{z},W_{z},c_{z},z)$ such that $k_z = \min_{v \in \sA} k_v$. The agents perform this task using pull operations as in the  standard \gossip\ broadcast protocol \cite{Shah09}, taking $O(\log n)$ rounds. More precisely, at every round, every agent $u$ stores      
 the  current  ``minimal'' certificate, i.e., that  with the  minimum value of $k_{z}$ he has seen so far and $u$ asks (via a pull operation)
  to  a random neighbor $v$ his  current minimal  certificate.  We call $\CE^{\min}_u$ the certificate owned by $u$ at the end of the \findmin\ phase. 
  
 \noindent
 -  The \coherence\ phase is performed in order to ensure that all the agents posses  the same certificate, namely the one resulting from the \findmin\ phase. In particular, agent $u$ sends his $\CE^{\min}_u$ to a logarithmic number of randomly chosen agents and he makes the protocol fail\footnote{For instance, the agent can enter in an invalid state by supporting a color not in $\Sigma$.} if he receives a different certificate $\CE^{\min}_v$ from an agent $v$. 

\noindent 
 - At the end of the \verification\ phase, every agent $u$ agrees on the color $c_{z}$ if the votes in $W_{z}$ are compatible with the votes in $L_u$. The votes are not compatible if there is a vote in $W_{z}$, say a vote given to $z$ by $w$, which is different from the vote to $z$ by $w$ stored in $L_u$ (hence, $u$ pulled $w$ in the \asking\ phase).

 \smallskip
   In     Subsection \ref{ssec:nonselfish}, we   show that the proposed protocol w.h.p. achieves fair consensus in presence of
 at most $\alpha n$ faulty agents, while, in Subsection \ref{ssec:selfish}, we prove that our protocol is a \equilibrium\ for any $t = o(n/\log n)$.

\subsection{Analysis of the protocol in the  cooperative setting} \label{ssec:nonselfish}


In this section we analyse Protocol $\prot$ when all the active agents follow $\prot$. We first give the concept of a \textit{good} execution of $\prot$. In a good execution, every active agent receives $\Theta(\log n)$ votes, all the $k_u$ values are all distinct (so, $k_{\min}$ is unique), and after the   \findmin\ phase, every active agent agrees on the same Certificate of minimal value. Formally, we introduce the following definition:

 
 

\begin{definition}\label{defgood}
\def \textbf{Good execution:} Let $q(\prot , \vec c) \in   Q(\prot , \vec c)$ be  a random execution of the protocol.
We say that
$q(\prot, \vec c)$ is \emph{good} (and define
 $ \good \subset \Omega$ as  the set of all good executions) if all the following events hold:

\begin{enumerate}[leftmargin=*]

\item Every agent in $\sA$ receives $\Theta(\log n)$ votes.
\item  The $k_u$ values are all  distinct (so, $k_{\min}$ is unique).
\item Let   $\CE^{min}$ be the certificate of the agent getting    the minimal  value $k_{\min}$.
Then, after the   \findmin\ phase,  for every active agent $u$, we have    $\CE_{u}^{min} = \CE^{min}$.

\end{enumerate}

\end{definition}

Lemma \ref{theogood} below shows that, if number of non-faulty agents is $\Theta(n)$, a random execution of $\prot$ is good w.h.p. 

\begin{lemma}\label{theogood}
Let $\ffaults$ be an absolute constant such that $0 \leq \ffaults <1$. If the number of faulty agents is at most $\ffaults n$, then the
   random execution of $\prot$ (with a suitable choice of parameter $\plenght = \plenght(\ffaults)$) is  good, w.h.p., i.e.\
 $\Pr_{\Omega}(\good) \geq 1 - \frac 1{n^{\Theta(1)}}$.
\end{lemma}

\skproof
We assume that there are at most $\ffaults n$ faulty nodes  and  that
all the active agents follow the algorithm for $\plenght \log n$ rounds (for a suitable  constant $\plenght(\ffaults)$).
As for  Point 1, for every agent $v$, consider the random variable $X_v$ that counts
  the number of votes   agent $v$ will get after the  \voting\ phase.   In this  phase, at each of the $\plenght \log n$ rounds,
every active node chooses independently  and u.a.r. one agent to vote. So,  $X_v$ can be written as
the sum of $\Theta(n\log n)$ mutually independent Bernoulli random variables.
   Using   Chernoff's bound (see Lemma \ref{lemma:cb})  on every    random variable $X_v$ and the Union Bound, we have that
      (for   a suitable choice of parameter $\plenght=\plenght(\ffaults)$)    two positive
constants $\beta_1, \beta_2$ exist such that
\[ \beta_1 \log n \leq  X_u \leq \beta_2 \log n \, ,  \ \forall u \in \sA, \, \mbox{w.h.p.}  \]

 \noindent
 As for Point 2, since  the $k_u$ values are independently chosen
     u.a.r. in $[m] = [n^3]$,  using standard argument, there will be
      no   collisions  and, thus,  the minimum of these values is unique, w.h.p.

\noindent
As for     Point 3, observe that  the \findmin\ phase is equivalent to a standard single-source
broadcast operation of the message $\CE^{min}$ on the complete
subgraph induced by the subset $\sA$ of active agents. The convergence time of this basic task on the complete graph
for the  \gossip\ model
- when agents use the pull mechanism - is known to be $\Theta(\log n)$ (w.h.p) \cite{Shah09}.
 The only difference here is the presence of faulty agents. However,  by a suitable choice of
 the constant $\plenght=\plenght(\ffaults)$, we can
 easily adapt the analysis in  \cite{Shah09} for the
  complete subgraph induced by any subset $\sA$ of active agents provided that
  $|\sA| \geq  (1-\alpha) n$ (essentially, the presence of $\alpha n$ faulty agents is balanced by a slightly
  longer broadcast phase).
       \qed
       
The three properties guaranteed by a good execution are the key ingredients in the proof of the next theorem stating that Protocol $\prot$ achieves a fair-consensus.

\begin{theorem}\label{proport}
Let $\ffaults$ be an    absolute constant such that $0 \leq \ffaults <1$.
 If the number of faulty agents is at most $\ffaults n$, Protocol $\prot$ (with a suitable choice of parameter $\plenght = \plenght(\ffaults)$)
  computes a fair consensus within $O(\log n)$ rounds and using messages of size $O(\log^2 n)$, w.h.p.

\end{theorem}

\skproof
Conditioning  to $q(\prot, \vec c) \in \good$ (an event that holds   w.h.p. because of   Lemma \ref{theogood}), we can
assume that the protocol does not fail. Indeed, from Definition \ref{defgood} (property 1), for every active agent $u$ the value
 $k_u$ is defined, from Definition \ref{defgood} (property 2) the value $k_{min}$ is unique and, from Definition \ref{defgood} (property 3), after  the \findmin\ phase all active agents converge to  a unique $\CE^{min}$. So, there are no multiple minimal
 certificates that can make fail the \coherence\ phase and in the \verification\ phase the Certificate is valid (each agent votes as declared in the \asking\ phase). By  simple  probabilistic arguments, the computation of $k_u = \sum_{h \in W_u} h \mod m$
 performed  by every agent $u$ implies that every agent has the same chance to
 get the (unique) minimal value.  So, the  protocol computes a fair leader election and
    the network converges to the leader  color in the \verification\ phase. The protocol terminates
     in  $O(\log n)$ rounds (by construction) and the largest message is the Certificate of the most
     voted agent that have size $O(\log^2 n)$. Indeed, thanks to  Definition \ref{defgood}.1, it gets
     $O(\log n)$ votes, each of them having   size $O(\log m) = O(\log n)$.
\qed


\smallskip
\subsection{Analysis of the protocol  in the presence of rational agents} \label{ssec:selfish}


In this section we analyse Protocol $\prot$ in the presence of rational agents and show that $\prot$ is a w.h.p.\ $t$-strong equilibrium, for any  $ t = o\left(\frac{n}{\log n} \right)$. We recall that $\sA$ is the set of active agents, $C$ the set of agents that deviate to a new set of local algorithms $\prot'_{C}$ while $\sA \setminus C$ is the subset of active agents that follow Algorithm \ref{alg}. W.l.o.g.\ we assume $C \subseteq \sA$. Moreover, we say that an agent $u$ is in the \emph{vote intention} of an agent $v \in \sA \setminus C$ if
 it holds  $(*,u) \in H_v$    (thus   in  the \voting\ phase $u$ will receive a vote of $v$). 
 
Following the same approach of Section \ref{ssec:nonselfish}, we first revise the notion of good execution in order to deal with the selfish behaviour of rational agents.


\begin{definition}\label{defgood2}
\def \textbf{Good execution:} Let $C$ be the coalition, $\prot'_{C}$ be the new set of local algorithms for $C$, and $q((\prot_{-C},\prot'_{C}), \vec c)$ be the random variable over $Q((\prot_{-C},\prot'_{C}), \vec c)$. We say that $q((\prot_{-C},\prot'_{C}), \vec c)$ is good (and define $\good' \subset \Omega'$ as the subset of all good executions) if   the following events hold:

\begin{enumerate}[leftmargin=*]
\item In the \asking\ phase each agent $u \in \sA$ receives
at least one pull request by an  agent   $v \in \sA \setminus C$ asking for (a copy of) $H_u$.
\item At the end of  the \coherence\ phase either  Protocol $(\prot_{-C},\prot'_{C})$ fails or every agent $v \in \sA \setminus C$ gets the same certificate $\CE^{\min}$.
\item At the end of the \asking\ phase, let $M \subset \sA$ be the set of agents that have received at least a pull request by an agent in $C$. Then for every agent $u \in \sA$, there exists an agent $v \in \sA \setminus (C \cup M)$ such that $u$ is in the vote intention of $v$ (i.e. $u$ receives the vote from $v$ in the \voting\ phase).
\end{enumerate}

\end{definition}

Under some reasonable assumptions of the number of faulty agents and the size of the coalition, the next lemma shows that the random execution of protocol is good w.h.p., even when a coalition deviates from $\prot$.

\begin{lemma}\label{theogood2}
Let $\ffaults$   be an   absolute constant such that $0 \leq \ffaults <1$.
 For any set of faulty agents of size at most $\ffaults n$ and for any coalition $C$ of size $ o\left(\frac{n}{\log n}\right)$, the random execution $q((\prot_{-C},\prot'_{C}), \vec c)$ (with a suitable choice of the parameter $\plenght = \plenght(\ffaults)$)
is  good, w.h.p.,  i.e. $\Pr_{\Omega'}(\good') \geq 1 - \frac 1{n^{\Theta(1)}}$.
\end{lemma}

\bigproof The theorem hypothesis imply that $\vert \sA \setminus C \vert = \ffaults'n$ for some constant $0 < \ffaults' < 1- \ffaults$ and we recall that each active agent in $\sA \setminus C$ runs each phase of  Algorithm \ref{alg}  for $\plenght \log n$ rounds.

\smallskip
\noindent
1) For any agent $u \in \sA$  and for any agent $v \in \sA \setminus C$, define the binary random variable $X_{u,v}=1$ iff in the \asking\ phase   agent $u \in \sA$ receives
a  pull request by $v$  asking for (a copy of) $H_u$. Since $v$ follows the protocol, it  easily holds that
\[ \Pr(X_{u,v}=0) = \left(1-\frac 1n\right)^{\plenght  \log n} \]
Since, all agents $v \in \sA \setminus C$ follow the protocol, thus making  mutually independent u.a.r. pull requests, we get

\[ \Pr(\forall v \in \sA \setminus C: \, X_{u,v}=0) = \left(1-\frac 1n\right )^{\ffaults' n \cdot \plenght  \log n}  \]
\[\leq e^{-\ffaults'  \plenght  \log n }  =
\frac1{n^{\ffaults'  \plenght}}\]
Finally, choosing a sufficiently large $\plenght$ and applying the Union Bound, we get
\[ \Pr(\exists u, \forall v \in \sA \setminus C : \, X_{u,v}=0) = \frac 1{n^{\Theta(1)}} .\]

\smallskip
\noindent
2) Assume that at the beginning of the \coherence\ phase there are at least two distinct Certificates, and let $\CE'$ be one of them. We thus consider the subset $X$ of $\sA \setminus C$ formed by all the agents having $\CE'$. Without loss of generality assume $\vert \sA \setminus (C\cup X) \vert \geq \vert X \vert$ thus $\vert \sA \setminus (C\cup X) \vert \geq \frac{\ffaults'}{2}n$. Then, using similar arguments to those in the proof of Point 1, we can   fix
$\plenght(\ffaults)$ such that  (after $\plenght \log n$ rounds) there is (at least) one  agent in $X$ that, following the protocol,
 will send his Certificate to an agent in $\sA \setminus (C\cup X)$ w.h.p. Then the protocol fails.

\smallskip
\noindent
3) Since $\vert C\vert = o(\frac{n}{\log n})$ and the length of the \asking\ phase is $O(\log n)$ rounds, the overall number of pull requests during this phase made by $C$ is $o(n)$. Thus we can assume that $\vert \sA \setminus (C \cup M)\vert \geq \lambda n$ for a fixed constant $\lambda$, with $0 \leq \lambda<\ffaults'$. Moreover, the overall number of votes sent by $\sA \setminus (C \cup M)$ (towards an agent chosen independently u.a.r. in $[n]$) in the \voting\ phase is greater than $\lambda\plenght n \log n$. Hence, since all the  agents in $\sA \setminus (C \cup M)$ follow the protocol,
using similar arguments to those in the proof of Point 1, we can   fix
$\plenght(\ffaults)$  in order to ensure that every agent in $\sA$ receives at least one vote from an agent in $\sA \setminus (C \cup M)$, w.h.p.
\qed

Lemma \ref{theogood2} ensures that, at the end of a good execution, the following facts hold w.h.p.:
\noindent
1) The vote intention of any agent $u$ in the coalition $C$ can be verified by at least one agent $v$ which does not belong to $C$;
\noindent
2) If the protocol does not fail, then all the agents which does not belong to $C$ agree on the same certificate $\CE^{min}$ and they check the same set of votes $W_{\min}$;
\noindent
3) All the agents receive at least one vote from an agent which does not belong to $C$. This guarantees that for every agent $u$ the value $k_u$ cannot be controlled by the coalition $C$, and thus, according to the protocol rules, $k_u$ is chosen u.a.r.\ in the range $[1,m]$; 
\noindent
The three facts above will be used to prove that the protocol $\prot$ is a \equilibrium.

\begin{theorem} \label{thm:main}
Let $\ffaults$   be  an   absolute constant such that $0 \leq \ffaults <1$.
For any set of faulty agents of size at most $\ffaults n$,
protocol $\prot$  (with a suitable choice of the parameter $\plenght = \plenght(\ffaults)$)    is   a \equilibrium\
for any  $ t = o\left(\frac{n}{\log n} \right)$.
\end{theorem}


\bigproof
Let us consider an arbitrary    coalition $C$ of at most $t$ agents and fix the set of local algorithms
$\prot_C'$ for them. We need the following  preliminary definitions:

\begin{itemize}[leftmargin=*]
\item $h^*_{v,u}$ is the vote declared by agent $v$ for the agent $u$ in the first declaration of $v$ to some
 agent $z \in \sA \setminus C$ during the \asking\ phase  (we recall that if $v$ has  not correctly replied to $z$ or $(*,u) \not \in H_v$ then $h^*_{v,u} = 0$).
 We also define $k^*_u = \sum_{v \in \sA} h^*_{v,u}$. Notice that $k^*_u$ may be different from $k_u$, the latter being
  the value that $u$ should declare
  (in the Certificate $\CE_u$)  during the \findmin\ phase and also observe that $k_u$ is a value agent $u$ can lie on. The difference between $k^*_u$ and $k_u$ leads us to introduce the following two concepts of winner.

\item We call \emph{Winner} the agent whose label is contained in the (unique - whenever  the  execution is good) certificate $\CE^{\min}$ after the \coherence\ phase. Notice that the certificate $\CE^{\min}$ contains the value $k_{min}$ which is the minimum value among the declared values $k_u$.

\item Let $a = \arg \min_{v \in \sA \setminus C} k_v$, and let $b=\arg \min_{v \in C} k^*_v$. We say that the \emph{Legitimate Winner} is $a$ if $k_a < k_b^*$, and it is  $b$ otherwise. Notice that the definition of \emph{Legitimate Winner} does not depend on the values $k_v$ declared by agents in $C$ and thus it may be case that \emph{Winner} and \emph{Legitimate Winner} are not the same. In particular we are interested in the following distinct two events.

\item For any $u \in \sA$, let $E_u$ be the event "the \emph{Legitimate Winner} is $u$" and
 define $E_C = \cup_{u \in C} E_u$. Furthermore $E'_C$ is the event ``the \emph{Winner} is an agent in $C$''.
\end{itemize}

We now prove that, in a non-failing good execution of $(\prot_{-C},\prot'_{C})$, if the \emph{Legitimate Winner} is not in $C$, then the \emph{Winner} is not in $C$ as well.

\begin{claim}\label{cl:1}
Let us consider any good execution which does not fail. Conditioning to the event $\bar E_C$,  the Winner turns out to be the Legitimate Winner (hence $\bar E_C$ implies $\bar E'_C$).
\end{claim}

\bigproof (of Claim \ref{cl:1})
The proof argument is by  contradiction.   Assume that the \emph{Legitimate Winner} $v$ belongs to $\sA  \setminus C$ and  the protocol does not converge to $c_v$. Then this happens only if agent $v$ accepts a certificate $\CE_u=(k_{u},W_{u},c_{u},u)$ different from his own Certificate $\CE_v=(k_{v},W_{v},c_{v},v)$ and such that $k_u < k_v$. Notice that, by definition of \emph{Legitimate Winner}, $u$ must belong to $C$ and $k_u \neq k^*_u$. Hence, thanks to Definition \ref{defgood2} (property 2),
at the end of  the \coherence\ phase every agent $v \in \sA \setminus C$ gets the same certificate $\CE_u$ (hence, the same set $W_u$) and thanks to Definition \ref{defgood2} (property 1),
 some agent $z \in \sA\setminus C$ exists whose local data  (i.e. $L_z$) is not consistent w.r.t.\ $W_u$  thus making the protocol fail. A contradiction.
\qed

\begin{claim} \label{cl:2}
Let us consider any good execution, every agent in $\sA$ has the same chance to be the Legitimate Winner (i.e.  $\forall u \in \sA$, $\Prob(E_u) = \frac{1}{\vert \sA \vert }$).
\end{claim}

\bigproof (of Claim \ref{cl:2})
We will argue that (i) for any $b \in C$ we have that $k^*_b$ is u.a.r. in $[m]$ and (ii) for any $a \in \sA \setminus C$ we have that $k_a$ is u.a.r. in $[m]$. To prove (i), notice that  $k_b^*$ is defined in the \asking\ phase: Thanks to Definition \ref{defgood2} (property 3) at the end of this phase, for every $b \in C$ there is still at least one agent $z$ in $\sA \setminus C$ that voted $b$ and was not pulled by any agent of $C$. Since $z \in \sA \setminus C$, $h^*_{z,b}$ coincides to the vote of $z$ actually given (in the \voting\ phase) to $b$, which is distributed u.a.r. in $[m]$. For the principle of deferred decision this implies that $k^*_b$ is u.a.r. in $[m]$ as well. To prove (ii), notice that  $k_a$ is determined in the \voting\ phase: Thanks to Definition \ref{defgood2} (property 3) at the end of the \voting\ phase, for every $a \in \sA \setminus C$ there is still at least one agent $z$ in $\sA \setminus C$ that voted for $a$ and was not pulled by any agent of $C$. Since $z \in \sA \setminus C$, $h^*_{z,a}$ coincides to the vote of $z$ actually given (in the \voting\ phase) to $a$ which is distributed u.a.r. in $[m]$. For the principle of deferred decision this implies that $k_a$ is u.a.r. in $[m]$ as well. Observe that, since both $z$ and $a$ are in $\sA \setminus C$, $h^*_{z,a}$ cannot be discovered by any agent in $C$ during the \voting\ phase.\\
Claim 2 follows from (i), (ii), and from the fact that, for simple symmetry argument, any agent has the same chance to get the minimal value.
\qed

Given any subset $X \subseteq \sA$ and any color $c \in \Sigma$, we define  $N(X,c)$  as the number of agents
  in $X$ supporting $c$. Then Claims \ref{cl:1} and \ref{cl:2}, easily imply the following properties of a good execution.

\begin{claim}\label{cl:3} Let us consider any good execution which does not fail. Conditioning to the event $\bar E_C$,
the protocol converges to a color $c \in \Sigma$
 with probability $\frac{N(\sA\setminus C,c)}{\vert \sA\setminus C \vert}$.
\end{claim}

\begin{claim} \label{cl:4}
Let us consider any good execution which does not fail.  Then it holds that
$\Pr(E'_C) \leq \frac{|C|}{|A|}$. \end{claim}

\noindent
Thanks to Lemma \ref{theogood2} we can now consider only good executions (i.e. $q((\prot_{-C},\prot'_{C}), \vec c) \in \good'$) where Claim \ref{cl:3} and Claim \ref{cl:4} do hold. We now use such claims to show by contradiction that the protocol is a \equilibrium. 

\noindent
For any  $c \in \Sigma$ define  $\Pr(c)$  as the probability the protocol converges to color $c$.  Then  for every agent $ u \in C$,
 we can evaluate  his  expected utility\footnote{Recall that all the events in $\Omega'$ and $\Omega$ are conditioned to $q((\prot_{-C},\prot'_{C}), \vec c) \in \good'$ and $q(\prot, \vec c) \in \good$, respectively. For the sake of clarity, with a little abuse of notation, we will not explicitly write it.}:
\begin{align}
& \Exp_{\Omega'}[r_u((\prot_{-C},\prot'_{C}), \vec c)] = \nonumber \\
& \Pr(f(q((\prot_{-C},\prot'_{C}), \vec c)=c_u) - \chi \Pr(f(q((\prot_{-C},\prot'_{C}), \vec c) = \perp) \leq \nonumber \\
& \Pr(f(q((\prot_{-C},\prot'_{C}), \vec c)=c_u) = \nonumber \\
& \Pr(E'_C)\Pr(c_u \vert E'_C)+ \Pr( \bar E'_C) \Pr(c_u \vert \bar E'_C) \nonumber
\end{align}
Hence,
\begin{align} \label{eq:exputility}
& \Exp_{\Omega'}[r_u((\prot_{-C},\prot'_{C}), \vec c)] \leq  \nonumber
\\
&   \Pr(E'_C)\Pr(c_u \vert E'_C)+ \Pr( \bar E'_C) \Pr(c_u \vert \bar E'_C)
   \end{align}
Thanks to Claim \ref{cl:3} and \ref{cl:4}, we can fix some $\delta \in [0,1]$ such that the above formula can be rewritten as:
\begin{align}
& \left( \frac{\vert C \vert}{ \vert \sA \vert} -\delta \right)\Pr(c_u \vert E'_C) + \left(\frac{\vert \sA \setminus C \vert}{\vert \sA \vert} + \delta \right) \frac{N(\sA\setminus C,c_u)}{\vert \sA \setminus C\vert} \nonumber
\end{align}
Note that, thanks to Theorem \ref{proport}, if all the agents follow Protocol $\prot$ the expected utility of every agent $u$ is
$N(\sA,c_u)/ \vert \sA \vert$. According to Definition \ref{tstrong}, in order to have a profitable deviation for $C$, for every agent $u \in C$, we should have:
\[
\Exp_{\Omega'}[r_u( (\prot_{-C},\prot'_{C}) , \vec c)] > \Exp_{\Omega}[r_u( \prot , \vec c)] \]
Thanks to Inequality \ref{eq:exputility}, the above condition implies that 
\begin{align*}
& \left(\frac{\vert C \vert}{ \vert \sA \vert} -\delta \right)\Pr(c_u \vert E'_C) + \left(\frac{\vert \sA \setminus C \vert}{\vert \sA \vert} + \delta \right) \frac{N(\sA\setminus C,c_u)}{\vert \sA \setminus C\vert} \\ & > \frac{N(\sA,c_u)}{\vert \sA \vert}
\end{align*}
Hence,
\begin{align*} 
& \Pr(c_u \vert E'_C) >   \\
& \frac 1{ \vert C \vert / \vert \sA \vert - \delta} \cdot \left(  \frac{N(\sA,c_u)}{\vert \sA \vert} - \left(\frac{\vert \sA \setminus C \vert}{\vert \sA \vert} + \delta \right)
\cdot \frac{N(\sA\setminus C,c_u)}{\vert \sA \setminus C\vert} \right) 
 \end{align*}
 It thus follows that 
\begin{align} \label{eq:prob1}
\Pr(c_u \vert E'_C) >  \frac{N(C,c_u) / \vert \sA \vert - \delta N(\sA\setminus C,c_u) / \vert \sA \setminus C \vert }{ \vert C \vert / \vert \sA \vert - \delta},
\end{align}
where in the last inequality we used the fact that, by definition, 
\[
N(C,c_u) = N(\sA,c_u) - N(\sA\setminus C,c_u)
\]

Let $\Sigma(C)$ be set of all the colors that are supported by at least one agent in $C$. So, for any $c \in \Sigma(C)$, 
Inequality \ref{eq:prob1} should hold. Then,  saturating the above inequalities over all colors in $\Sigma(C)$, we get:
\begin{align} \label{eq:satur}
 & \sum_{c \in \Sigma(C)} \Pr(c \vert E'_C) > \nonumber\\
& \sum_{c \in \Sigma(C)} \frac{N(C,c) / \vert \sA \vert - \delta N(\sA\setminus C,c) / \vert \sA \setminus C \vert }{ \vert C \vert / \vert \sA \vert - \delta}  
\end{align}
Since 
\[ \sum_{c \in \Sigma(C) } N(C,c) =  \vert C \vert, \]
then the r.h.s. of     Inequality \ref{eq:satur} can be rewritten as
 
\begin{align} \label{eq:sumofC}
 &\sum_{c \in \Sigma(C)} \frac{N(C,c) / \vert \sA \vert - \delta N(\sA\setminus C,c) / \vert \sA \setminus C \vert }{ \vert C \vert / \vert \sA \vert - \delta} = \nonumber 
 \\
&\frac{ \vert C \vert / \vert \sA \vert - \delta \sum_{c \in \Sigma(C)} N(\sA\setminus C,c) / \vert \sA \setminus C \vert}{ \vert C \vert / \vert \sA \vert - \delta} 
\end{align}
  Since, by definition, it holds that  
  \[
  \sum_{c \in \Sigma(C)} N(\sA\setminus C,c)  \leq  \vert \sA \setminus C \vert 
  \]
then we get 
\begin{align} \label{eq:sumofCa}
 & \frac{ \vert C \vert / \vert \sA \vert - \delta \sum_{c \in \Sigma(C)} N(\sA\setminus C,c) / \vert \sA \setminus C \vert}{ \vert C \vert / \vert \sA \vert - \delta} \geq 1 
\end{align}

From Inequalities \ref{eq:satur}-\ref{eq:sumofCa}, we should have $\sum_{c \in \Sigma(C)} \Pr(c \vert E'_C) > 1 $: This is clearly false.
Thus, there must be at least one agent in $C$ that will not increase his expected utility, concluding the proof.
\qed




\subsection{Useful probability bounds}
\begin{lemma}[Chernoff bounds]\label{lemma:cb}
Let $X = \sum_{i=1}^n X_i$ where $X_i$'s are independent Bernoulli random variables and let $\mu = \Expec{}{X}$. Then,
\begin{enumerate}[leftmargin=*]
\item For any $0 < \delta \leqslant 4$, $\Pr(X > (1 + \delta)\mu) < e^{-\frac{\delta^2\mu}{4}}$;
\item For any $\delta \geqslant 4$, $\Pr(X > (1 + \delta)\mu) < e^{-\delta\mu}$;
\item For any $\lambda > 0$, $\Pr(X \geqslant \mu + \lambda) \leqslant e^{-2 \lambda^2 / n}$.
\end{enumerate}
\end{lemma}

\section{Conclusions}
Efficient algorithmic methods for consensus tasks in fully-decentralized  systems where agents may reveal  a selfish behaviour is
a central issue that arises in several  scientific fields such as social networks, peer-to-peer networks, biological systems, e-commerce, and
crypto-currency. Hence,  the definition  of reasonable distributed models and   specific problems capturing some of the major  technical
questions is a  line of research that is currently attracting increasing interest from the distributed computing community.
One of the    technical goals in this context  is that of reducing
   \emph{local memory and communication cost} of the proposed consensus protocols.
Considering the specific network scenarios where  this kind of rational consensus may play an important role,
we believe this is an important question which is still far to be  well-understood. Our contribution provides a first
step for this general aim since it shows that, on complete networks, fair rational consensus can be obtained in logarithmic time in a communication model, the \gossip\ one,
 that severely restricts both local memory and message communication.

In our opinion,  two specific open     problems  ``suggested'' by our work  look rather interesting. The first one is to provide \gossip\ algorithms for rational fair consensus
 in other relevant classes of graphs, while the second one is the study of  this problem in the asynchronous (i.e. sequential)  \gossip\ model where, at every round,
 only one (possibly random) agent is awake.

\bibliography{long}{}

\begin{thebibliography}{10}

\bibitem{AbrahamDGH06}
I.~Abraham, D.~Dolev, R.~Gonen, and J.Y. Halpern.
\newblock Distributed computing meets game theory: robust mechanisms for
  rational secret sharing and multiparty computation.
\newblock In Eric Ruppert and Dahlia Malkhi, editors, {\em Proceedings of the
  Twenty-Fifth Annual {ACM} Symposium on Principles of Distributed Computing,
  {PODC} 2006, Denver, CO, USA, July 23-26, 2006}, pages 53--62. {ACM}, 2006.

\bibitem{AbrahamDH13}
I.~Abraham, D.~Dolev, and J.Y. Halpern.
\newblock Distributed protocols for leader election: {A} game-theoretic
  perspective.
\newblock In Yehuda Afek, editor, {\em Distributed Computing - 27th
  International Symposium, {DISC} 2013, Jerusalem, Israel, October 14-18, 2013.
  Proceedings}, volume 8205 of {\em Lecture Notes in Computer Science}, pages
  61--75. Springer, 2013.

\bibitem{AfekGFS14}
Y.~Afek, Y.~Ginzberg, S.L. Feibish, and M.~Sulamy.
\newblock Distributed computing building blocks for rational agents.
\newblock In Magn{\'{u}}s~M. Halld{\'{o}}rsson and Shlomi Dolev, editors, {\em
  {ACM} Symposium on Principles of Distributed Computing, {PODC} '14, Paris,
  France, July 15-18, 2014}, pages 406--415. {ACM}, 2014.

\bibitem{AFM07}
Nir Andelman, Michal Feldman, and Yishay Mansour.
\newblock Strong price of anarchy.
\newblock In {\em Proceedings of the Eighteenth Annual ACM-SIAM Symposium on
  Discrete Algorithms}, SODA '07, pages 189--198, Philadelphia, PA, USA, 2007.
  Society for Industrial and Applied Mathematics.

\bibitem{AAE07}
Dana Angluin, James Aspnes, and David Eisenstat.
\newblock {A Simple Population Protocol for Fast Robust Approximate Majority}.
\newblock {\em Distributed Computing}, 21(2):87--102, 2008.
\newblock (Preliminary version in DISC'07).

\bibitem{BCNPS15}
L.~Becchetti, A.~Clementi, E.~Natale, F.~Pasquale, and R.~Silvestri.
\newblock {Plurality Consensus in the Gossip Model}.
\newblock In {\em Proc. of the 26th Ann. ACM-SIAM Symp. on Discrete Algorithms
  (SODA'15)}, pages 371--390. SIAM, 2015.

\bibitem{BeiCZ12}
X.~Bei, W.~Chen, and J.~Zhang.
\newblock Distributed consensus resilient to both crash failures and strategic.
\newblock In {\em http://arxiv.org/abs/1203.4324; version 3}, 2012.

\bibitem{HHKM12}
Keren Censor-Hillel, Bernhard Haeupler, Jonathan Kelner, and Petar Maymounkov.
\newblock Global computation in a poorly connected world: Fast rumor spreading
  with no dependence on conductance.
\newblock In {\em Proceedings of the Forty-fourth Annual ACM Symposium on
  Theory of Computing}, STOC '12, pages 961--970, New York, NY, USA, 2012. ACM.

\bibitem{DGHILSSST87}
Alan Demers, Dan Greene, Carl Hauser, Wes Irish, John Larson, Scott Shenker,
  Howard Sturgis, Dan Swinehart, and Doug Terry.
\newblock {Epidemic algorithms for replicated database maintenance}.
\newblock In {\em Proc. of the 6th Ann. ACM Symposium on Principles of
  Distributed Computing (PODC'12)}, pages 1--12. ACM, 1987.

\bibitem{DGMSS11}
B.~Doerr, L.~A. Goldberg, L.~Minder, T.~Sauerwald, and C.~Scheideler.
\newblock {Stabilizing consensus with the power of two choices}.
\newblock In {\em Proc. of the 23rd Ann. ACM Symp. on Parallelism in Algorithms
  and Architectures (SPAA'11)}, pages 149--158. ACM, 2011.

\bibitem{GH05}
Andrew~V. Goldberg and Jason~D. Hartline.
\newblock Collusion-resistant mechanisms for single-parameter agents.
\newblock In {\em Proceedings of the Sixteenth Annual ACM-SIAM Symposium on
  Discrete Algorithms}, SODA '05, pages 620--629, Philadelphia, PA, USA, 2005.
  Society for Industrial and Applied Mathematics.

\bibitem{GroceKTZ12}
A.~Groce, J.~Katz, Thiruvengadam, and V.~Zikas.
\newblock Byzantine agreement with a rational adversary.
\newblock In {\em Proc. 39th ICALP, LNCS}, pages 561--572, 2012.

\bibitem{HT04}
Joseph Halpern and Vanessa Teague.
\newblock Rational secret sharing and multiparty computation.
\newblock In {\em Proceedings of the thirty-sixth annual ACM symposium on
  Theory of computing}, pages 623--632. ACM, 2004.

\bibitem{HV16}
J.Y. Halpern and X.~Vilaca.
\newblock Rational consensus: Extended abstract.
\newblock In {\em Proc. ACM PODC'16}, pages 561--572, 2016.

\bibitem{HP02}
Yehuda Hassin and David Peleg.
\newblock Distributed probabilistic polling and applications to proportionate
  agreement.
\newblock {\em Inf. Comput.}, 171(2):248--268, January 2002.

\bibitem{KSSV00}
Richard Karp, Christian Schindelhauer, Scott Shenker, and Berthold Vocking.
\newblock {Randomized rumor spreading}.
\newblock In {\em Proc. of the 41th Ann. IEEE Symp. on Foundations of Computer
  Science (FOCS'00)}, pages 565--574. IEEE, 2000.

\bibitem{KDG03}
David Kempe, Alin Dobra, and Johannes Gehrke.
\newblock {Gossip-Based Computation of Aggregate Information}.
\newblock In {\em Proc. of 43rd Ann. IEEE Symp. on Foundations of Computer
  Science (FOCS'03)}, pages 482--491. IEEE, 2003.

\bibitem{bit_crypto}
Arvind Narayanan, Joseph Bonneau, Edward Felten, Andrew Miller, and Steven
  Goldfeder.
\newblock {\em Bitcoin and Cryptocurrency Technologies}.
\newblock Princeton University Press, 2015.

\bibitem{Shah09}
Devavrat Shah.
\newblock Gossip algorithms.
\newblock {\em Found. Trends Netw.}, 3(1):1--125, January 2009.

\end{thebibliography}
\bibliographystyle{plain}

\end{document}